\pdfoutput=1

\documentclass{llncs}
  \usepackage{article}
  \usepackage{rev}
  \usepackage[draft=true]{minted}
  \usepackage{booktabs}

  \usepackage{dcolumn}
  \newcolumntype{d}[1]{D{.}{.}{#1} }

  \usepackage[inline]{enumitem}
  \setlistdepth{10}
  \newlist{steps}{itemize}{9}
  \setlist[steps,1]{label={-}}
  \setlist[steps,2]{label={-}}
  \setlist[steps,3]{label={-}}
  \setlist[steps,4]{label={-}}
  \setlist[steps,5]{label={-}}
  \setlist[steps,6]{label={-}}
  \setlist[steps,7]{label={-}}
  \setlist[steps,8]{label={-}}
  \setlist[steps,9]{label={-}}

  \usepackage{siunitx}

\newcommand{\cc}{{\kern-.1em}\texttt{:}{\kern-.25em}\texttt{:}}
\newcommand{\libvcsn}{\texttt{vcsn}\cc\xspace}
\newcommand{\libVcsn}{\texttt{Vcsn}\cc\xspace}
\newcommand{\libdyn}{\texttt{dyn}\cc\xspace}

\newminted[cxxcode]{cpp}{fontsize=\footnotesize}
\newmintinline[cxx]{cpp}{fontsize=\footnotesize}
\newmintinline[pyinline]{python}{fontsize=\footnotesize}
\newminted[plain]{text}{fontsize=\footnotesize}

\usepackage{vcsn}
\newcommand{\vcsnu}{\textsc{Vaucanson}~1\xspace}
\renewcommand{\eword}{\varepsilon}

\renewcommand{\var}[1]{\textit{\sffamily #1}}

\usepackage[numbers]{natbib}

\title{Runtime Template Instantiation for \Cxx}
\author{\XX{Akim Demaille}\\{\scriptsize \SvnDate \SvnRev}}
\institute{%
  \XX{LRDE, EPITA, \email{akim@lrde.epita.fr}}
}

\begin{document}
\maketitle

\begin{abstract}
  Performance, genericity and flexibility are three valuable qualities for
  scientific environments that tend to be antagonistic.  \Cxx provides
  excellent support for both performances and genericity thanks to its
  support for (class and function) templates.  However, a \Cxx templated
  library can hardly be qualified as flexible: data of unexpected types
  cannot enter the system, which hinders user interactions.  This paper
  describes the approach that was taken in the \vcsn platform to add
  flexibility on top of \Cxx templates, including runtime template
  instantiation.
\end{abstract}

\section{Introduction}

\subsection{Performance, Genericity, Flexibility}
In scientific environments dealing with large data (image or signal
processing, graphs, computational linguistics\ldots), performance,
genericity and flexibility are three valuable qualities that tend to be
antagonistic.

By \emph{performance}, we mean the efficiency of both the data structures
and the algorithms.  By \emph{genericity}, we mean the means to apply these
structures and algorithms to a wide set of basic data types.  By
\emph{flexibility}, we mean the ability to support new types at runtime, to
support seamless interaction with the user, loading files, etc.

\Cxx excels at blending performance and genericity thanks to its excellent
support of compile time code generation: templates.  However, precisely
because code generation is at compile time only, \Cxx is not flexible: one
cannot write a \Cxx program that would load a standard list
(\cxx{std::list}) for a type for which support was not compiled in.  The
world is \emph{closed} at compile-time: new types cannot be introduced.

Actually, \Cxx provides some flexibility when traditional \ac{oop} is used,
with \emph{virtual} member functions, but it is known that in this case the
performances are degraded.  At the other end, so called \emph{dynamic
  languages} such as Python offer excellent flexibility, and genericity:
thanks to their dynamic typing, in interactive sessions, users can easily
call algorithms on newly created data type, serialize data, etc.  However,
performances are poor.

This opposition shows in scientific libraries.  For instance, consider the
domain of graphs/networks.  At one end lie heavily templated \Cxx libraries
such as the \ac{bgl}~\cite{siek.01.bgl} which offers unparalleled genericity
(at the expense of being hard to tame) and excellent performances.  However,
because it relies on templates, it falls short on flexibility.

On the other hand of the spectrum, Networkx~\citep{hagberg.2008.scipy} also
offers a wide range of algorithms, and great flexibility, because it is
entirely written in Python\footnote{%
  However in some cases it uses NumPy and SciPy, Python libraries that are
  partially written in C/\Cxx/Fortran for performances.
}.  This comes at a cost: performance penalties such as 20x for
single-source shortest path for instance, and sometimes much higher
\citep{peixoto.2014.gtp}.

Somewhere in the middle, Graph-tool~\citep{peixoto.2014.graph-tool} offers
the efficiency of \Cxx and the user friendliness of Python: its core is a
Python binding of \ac{bgl}.  However this is at the expense of genericity: a
finite set of parameters were chosen for \ac{bgl}, and only this type of
graphs is supported.

The domain of automata theory presents a similar pattern.  A library such as
OpenFst offers genericity (it supports a large range of automaton types) and
a remarkable efficiency~\citep{allauzen.07.ciaa}.  It is a templated \Cxx
library.  Conversely, the FAdo environment~\citep{almeida.2009.ciaa} is
flexible, very user friendly, but offers poor performances.  It is written
in Python, with bits of C where Python is too slow.

\medskip

This paper describes the approach that was taken in the \vcsn platform to
reconcile performance, genericity and flexibility in
\Cxx.  We claim that this approach can be applied in various domains, under
some conditions exposed further.  However, because it was applied only once,
and because we believe it is simpler exposed on a specific domain, this
paper will focus on automata.

\subsection{A Few Notions of Automata Theory}

\begin{figure}[t]
  \centering
  \includegraphics{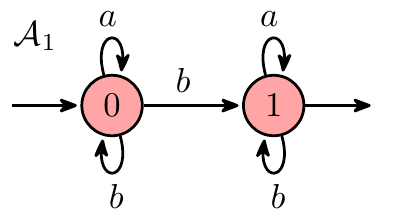}
  \hspace{1cm}
  \includegraphics{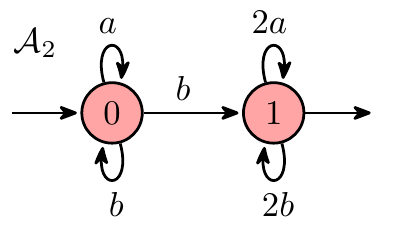}
  \caption{Automaton $\Ac_1$ is labeled by letters in $\{a, b\}$.  Its
    states are 0 and 1.  State 0 is initial, and State 1 is final.  It
    features five transitions:
    $\{(0, a, 0), (0, b, 0), (0, b, 1), (1, a, 1), (1, b, 1)\}$.  It accepts
    words on $\{a, b\}$ with at least one $b$.  The labels of Automaton
    $\Ac_2$ are letters in $\{a, b\}$, its weights are in $\mathbb{Z}$.
    Implicit weights denote the neutral of the multiplication: $b$ means
    $1b$. Its \emph{context} is $C_2 = \{a, b\} \rightarrow \mathbb{Z}$.}
  \label{fig:automaton}
\end{figure}

Finite automata are well studied (and commonly taught) devices that
recognize languages, i.e., accept or refuse words.  They are composed of a
finite number of \emph{states}, some of them being \emph{initial} and/or
\emph{final}, and a set of \emph{transitions}: triples
$(\mathit{source}, \mathit{label}, \mathit{destination})$, where
$\mathit{source}$ and $\mathit{destination}$ are states, and
$\mathit{label}$ is a letter.  $\Ac_1$ in \cref{fig:automaton} is a simple
automaton.  A word $u$ is \dfn{accepted} by an automaton if there exists a
path from an initial state to a final state that follows transitions whose
concatenated labels equal $u$. $\Ac_1$ accepts $bb$, and refuses $aa$.

Actually, the theoretical foundations are more
general~\citep{sakarovitch.09.eat}.  Labels may be chosen to be letters or
$\eword$ (the \dfn{empty word}, neutral for the concatenation), or words, or
even tuples of labels (for instance to model \dfn{transducers}: automata
with input \emph{and} output).  However, labels must belong to a
\emph{monoid}: they can be concatenated, concatenation is associative and
has a neutral element.

A transition is \emph{spontaneous} if it is labeled by $\eword$.  An
automaton without spontaneous transitions is \emph{proper}.  $\Ac_1$ is
proper.

Automata can be equipped with \emph{weights} (also called
\emph{multiplicities}): evaluating a word on an automaton gives more than
yes/no, it provides a ``score'' (e.g., a probability).  Weights along a path
are multiplied, and weights between paths are added.  For instance $\Ac_2$
evaluates $bb$ as 3: path $(0, 0, 1)$ computes $1 \times 1$, and path
$(0, 1, 1)$ computes $1 \times 2$.  In fact $\Ac_2$ computes the value of
binary numbers written on $\{a, b\}$.  Mathematically, the set of weights is
said to be a \emph{semiring}: a structure $\bra{\K, +, \cdot, \zeK, \unK}$
such that $\bra{\K, +, \zeK}$ is a commutative monoid,
$\bra{\K, \cdot, \unK}$ is a monoid, $\cdot$ distributes over $+$, and
$\zeK$ is an \dfn{annihilator} for~$\cdot$:
$\forall k \in \K, k \cdot \zeK = \zeK = \zeK \cdot k$.

\subsection{Outline}

\paragraph{Contributions.}
This paper presents a way to add dynamic polymorphism on a templated \Cxx
library.  This dispatching does not require arguments to be members of an OO
hierarchy of classes; in a way, it lifts \Cxx static overloading and
function templates into runtime polymorphism.  Flexibility, and in
particular support for an open world of types, is implemented using
introspection, runtime code-generation, compilation and loading.  We also
introduce Value/ValueSet, a design that allows an efficient implementation
of algebraic entities such as monoids, semirings, etc.

While this paper focuses on automata, the techniques may be applied in other
domains where large structures depend on many small values, for instance
image processing.  However because we are particularly interested in adding
runtime instantiation for templates, a
\Cxx-specific feature, and because the implementation uses so many \Cxx
features, we do not think exposing our approach independently from this
language would make any sense.  We tried to stay away from technical issues,
but it is quite impossible to do so without hiding criticial components.
Hence, the reader is expected to be comfortable with modern \Cxx.

\medskip

\paragraph{Outline.}
In \cref{sec:vcsn::} we describe \libvcsn, the base layer, a templated \Cxx
library, and in particular the Value/ValueSet design.  \cref{sec:dyn::}
details how the second layer, \libdyn, adds some runtime introspection,
runtime polymorphism, and runtime instantiation.  In \cref{sec:discussion}
we discuss some of the ideas and techniques that were used, what they make
possible, and some benchmarks.  \cref{sec:works} is dedicated to previous
and future works.  \cref{sec:conclusion} concludes.

\section{The Template-Based Library: \textsc{vcsn::}}
\label{sec:vcsn::}

The bottom layer of \vcsn is \libvcsn, a heavily templated \Cxx library.  It
consists of a set of data structures ---following a design which we call
\emph{Value/ValueSet}--- and algorithms on these data structures.

\subsection{Value/ValueSet}

\libVcsn deals with a wide range of value types: from labels and weights to
automata, \dfn{rational expressions} (aka \dfn{regular expressions}) and
others.  Labels and weights are typically small (Booleans, bytes, integers,
etc.), while automata and expressions are large, possibly very large
(hundreds of thousands of states).

A single \Cxx type may actually correspond to several algebraic structures.
For instance, \cxx{bool} may model either $\B$, the traditional Booleans
(where the addition is ``or'', hence $1 + 1 = 1$), or the field $\F_2$ which
corresponds to arithmetic modulo 2 (i.e., the addition is ``xor'':
$1 + 1 = 0$).  As another example, \cxx{int} may be equipped with the
traditional addition and multiplication, or $\min$ as an addition
(\cxx{MAX_INT} is its neutral) and $+$ as multiplication (0 is its
neutral)\footnote{%
  If $\Ac_2$ from \cref{fig:automaton} is interpreted with $\min$ and $+$,
  then $bb$ is evaluated as $0 = \min(0 + 0, 0 + 2)$.  More generally,
  $\Ac_2$ returns twice the number of $a$ after the last $b$, or 0 if there
  is no $b$.
}.

Since native data types such as \cxx{bool} or \cxx{int} lead to several
possible algebraic structures, they don't suffice to model our weights.  The
traditional design would then introduce classes such as
\cxx{traditional_boolean} with an attribute of type \cxx{bool}, and another
\cxx{f2_integer} also with an attribute of type \cxx{bool}.  However, this
design does not scale in the case of stateful valuesets.

Consider the case of labels: they are always defined with respect to a
specific alphabet; labels with invalid letters must be rejected.  If we were
to apply the straightforward design, the constructor of the labels would
need an access to the alphabet, and possibly some other runtime parameters.
Hence, a label as simple as a \cxx{char} would require an additional 4x or
8x payload penalty to carry around at least a pointer to its alphabet.

To address this issue we use the Value/ValueSet design, that goes somewhat
backwards compared to \ac{oop}: traditional objects are split in value on
the one hand, and operations on the other hand.  The \emph{Values} (such as
\cxx{bool} or \cxx{int}) are ``dumb'': they do not provide any operation,
they cannot even print themselves.  The \emph{ValueSets} (such as \cxx{b}
for $\B$ and \cxx{f2} for $\F_2$, \cxx{z} for $\Z$ and \cxx{zmin} for $\Z$
with $\min$ and $+$) provide all the operations: construction (including
validation), conversion, addition, multiplication, access to specific values
(the neutrals), etc.

\subsection{Concepts of \libvcsn}

In our design, types of automata are captured by two sets: the
\dfn{labelset} is the set of the valid labels (e.g., $\{a, b\}$: only
letters, or $\{a, b\}^*$: words of any length on $\{a, b\}$), and the
\dfn{weightset} is the set of the valid weights (e.g., $\mathbb{B}$ for
$\Ac_1$ and $\mathbb{Z}$ for $\Ac_2$).  A pair composed of a labelset $L$
and a weightset $W$ is named a \emph{context}, and is denoted
$L \rightarrow W$.

ValueSets in \libvcsn comply with a handful of concepts: WeightSet,
LabelSet, ExpressionSet for rational expressions, PolynomialSet etc.  We
also provide a \cxx{tupleset} variadic class template that allows define
Cartesian products of ValueSets.  For instance, $\F_2 \times \Z_\text{min}$
is a valid WeightSet, and $\{a, b\} \times \{x, y, z\}^*$ models labels that
are pairs whose first member is either an $a$ or a $b$, and whose second
member is any string of $x, y$ and $z$.

\Cxx features an extremely powerful model of templates: classes and
functions can depend on formal \emph{template parameters}: compile-time
meta-variables denoting types or values.  Obviously a given template expects
its actual template parameters to comply with some constraints, such as
``supports an addition'', ``is default constructible'' and so forth.
Introduced by \ac{stl}, the \Cxx community names \emph{concepts} these
requirements on template parameters.  Unfortunately \Cxx does not support
concepts, yet.  We will follow the syntax of the latest proposal for concept
in \Cxx~\citep{sutton.2013.3701,sutton.2014.n4040}.

The \dfn{LabelSet}, the set of valid labels, must be a subset of a monoid;
for instance, given a letter type \cxx{A}, \cxx{letterset<A>} denotes the
set of single-letter labels, and \cxx{wordset<A>} that of strings.

The theory requires weights to form a semiring.  The corresponding concept
in \libvcsn is WeightSet:
\begin{cxxcode}
template <typename T>
concept bool WeightSet() {
  using value_t = typename T::value_t;
  return requires(value_t a, value_t b) {
     equal_to(a, b) -> bool;      less(a, b) -> bool;
     zero()         -> value_t;   one()      -> value_t;
     add(a, b)      -> value_t;   mul(a, b)  -> value_t;
     // ...
  };
}
\end{cxxcode}

The class template \cxx{context} aggregates a LabelSet and a WeightSet.

\subsection{Conversions and Join}

There exists a subtype relation between labelsets~\citep{demaille.14.ciaa},
denoted $\subtype$.  For instance, since a letter can be seen as a string,
$\cxx{letterset<A>} \subtype \cxx{wordset<A>}$.  WeightSets also feature a
subtype relation.  For instance $\N \subtype \Z \subtype \Q \subtype \R$.

\libVcsn implements conversion routines to supertypes, for instance, to
convert an automaton of type $\{a, b\} \rightarrow \N$ to an automaton of
type $\{a, b, c\}^* \rightarrow \Q$.

As a more complex example, consider $\RatE{\{a, b\} \rightarrow \Q}$, the
set of rational expressions on letters $\{a, b\}$ with weights in $\Q$.
Rational expressions with weights in $\Z$ can be converted as expressions
with weights in $\Q$ (e.g., $3a^* \Rightarrow \frac{3}{1}a^*$), so
$\RatE{\{a, b\} \rightarrow \B} \subtype \RatE{\{a, b\} \rightarrow \Q}$.
Fractions can also been seen as expressions whose weights are fractions,
e.g, $\frac{1}{3} \Rightarrow \frac{1}{3}\eword$; hence
$\Q \subtype \RatE{\{a, b\} \rightarrow \Q}$.

\medskip

\libVcsn also provides routines to compute the smallest common
supertype of two types, which we call the \emph{join}.  For instance,
$\RatE{\{a, b\} \rightarrow \Q}$ is the join of
$\RatE{\{a, b\} \rightarrow \B}$ and $\Q$.

\subsection{The Algorithms}
\label{sec:algos}
With data structures on the one hand, and algorithms on the other hand,
\libvcsn is implemented like \ac{stl}.  Indeed, all the algorithms are
templated, yet expect their arguments to implement the needed concepts.
Many algorithms have a rather straightforward interface.  For instance
checking whether an automaton is proper (has no spontaneous transitions), or
evaluating a word:
\begin{cxxcode}
template <typename Aut>
auto is_proper(const Aut& a) -> bool;

template <typename Aut>
auto evaluate(const Aut& a, const word_t_of<Aut>& w) -> weight_t_of<Aut>;
\end{cxxcode}
where \cxx{word_t_of<Aut>} is the type of words corresponding to automata of
type \cxx{Aut}, and likewise for \cxx{weight_t_of<Aut>}.

\paragraph{Algorithms with Alternatives.}
Some computations can be implemented by different algorithms, with different
preconditions.  For instance minimization (which consists in producing a
smaller automaton) currently comes in three flavors:

\begin{cxxcode}
// a must be deterministic, its WeightSet must be Boolean.
template <typename Aut>
auto minimize_moore(const Aut& a) -> subset_automaton<Aut>;

template <typename Aut>
auto minimize_signature(const Aut& a) -> subset_automaton<Aut>;

// The LabelSet of a must be free.
template <typename Aut>
auto minimize_brzozowski(const Aut& a) -> brzozowski_automaton<Aut>;
\end{cxxcode}
where \cxx{subset_automaton} is a specific type of automaton whose
states ``remember'' the states of the input automaton they correspond to.
The \cxx{brzozowski_automaton} also maintains a connection with its
input automaton, but the relationship is different.

\medskip


\paragraph{Value Parameters.}
Some operations are templated by integers.  Consider for instance an
automaton of type $\{a, b\} \times \{x, y, z\}^* \rightarrow \Q$: it has two
tapes, one labeled with letters in $\{a, b\}$, and the second with words on
$\{x, y, z\}$.  The \cxx{focus} algorithm hides all the tapes except one.
The type of the resulting automaton ($\{a, b\} \rightarrow \Q$ or
$\{x, y, z\}^* \rightarrow \Q$) therefore depends \emph{statically} on the
tape number: it is a (compile-time) template parameter, not a (runtime)
function argument:

\begin{cxxcode}
template <unsigned Tape, typename Aut>
auto focus(const Aut& aut) -> focus_automaton<Tape, Aut>;
\end{cxxcode}

\paragraph{Variadic Functions.}
The synchronized product of automata is a binary, associative, operation.
What it computes is irrelevant to this paper.  However, much alike matrices
addition, when chained, it is much more efficiently handled as a single
variadic operation rather that repeated binary ones.  \libVcsn offers
the product as a variadic function: it accepts any number of automata,
besides, of heterogeneous type:
\begin{cxxcode}
template <typename... Auts>
auto product(const Auts&... as) -> product_automaton<Auts...>;
\end{cxxcode}

\medskip

To summarize, \libvcsn features algorithms that are alike but with
different return types (minimization), algorithms whose return type is
computed from those of the arguments (union), algorithms that statically
depend on integers (focus), and variadic algorithms (product).

\section{The Dynamically-Typed Library: \textsc{dyn::}}
\label{sec:dyn::}

The \libvcsn library presented in \cref{sec:vcsn::} is a typical instance of
a \Cxx template library: it is generic and efficient (it is on par with
OpenFst~\citep{demaille.13.ciaa}).  But it falls short on the flexibility:
if an object of an unexpected type is needed (e.g., an automaton is loaded
from a file), the library throws an exception and the program terminates.

The purpose of \libdyn is to bring flexibility \emph{on top of} \libvcsn.
This is achieved via type-erasure techniques, and some introspection support
to provide rich runtime-type information ---not to be confused with the \Cxx
native \ac{rtti} support.

\subsection{Introspection: Signatures}

To implement dynamic polymorphism we provide each type lifted in
\libdyn with a unique key, its \dfn{sname}.  Snames are
strings that look like \Cxx types, e.g. \cxx{"int"} for \cxx{int} arguments,
\cxx{"const std::string"} for \cxx{const std::string} etc.  This is a lot
like what the \Cxx's native \ac{rtti} provides via \cxx{typeinfo::name()};
however there is no guarantee on what \cxx{typeinfo::name()} actually
returns, and anyway, in some cases we need to ``lie'' to get powerful
interfaces between \libdyn and \libvcsn (see
\cref{sec:dyn::variations}).

Because snames are used as keys in tables, they are \dfn{internalized},
i.e., we use Boost.Flyweight~\citep{boost.www}, an implementation of the
Flyweight design pattern~\citep{gamma.95.dp}, to keep a single instance of
each value.  The main concern is not saving space, but having fast core
operations: since every value is represented by a unique instance,
comparisons and hashing are shallow as they are performed on the addresses.

The \cxx{sname()} function is a type traits~\citep{myers.95.cppr}, a
classical \Cxx technique which allows to cope with pre-existing types that
might not even be classes (e.g., \cxx{int}).

The Values do not need an sname, and actually, they cannot provide one:
knowing that a weight is of type \cxx{bool} does not tell anything about its
algebraic nature ($\B$ or $\F_2$?).  On the other hand,
ValueSets have an sname (e.g, \cxx{"b"}, \cxx{"f2"}, or
\cxx{"letterset<char_letters>"}).

Note that \cxx{sname()}, being based on traits, does not take any runtime
function argument, it depends solely on its static template parameters.
This is a key property for the registration
process~(\cref{sec:dyn::filling}).

\subsection{Type Erasure}
\label{sec:dyn:erasure}
The \libdyn layer relies on type erasure to store a \libvcsn pair
Value/ValueSet in a \cxx{dyn::value} object:\footnote{In order to keep the
  code snippet short and legible, we used the \Cxx{14} syntax which allows
  to leave the return type unspecified, to be deduced by the compiler.}

\begin{cxxcode}
namespace dyn {
  // Type-erased Value/ValueSet.
  template <typename Tag>
  class value { public:
    template <typename ValueSet>
    value(const ValueSet& vs, const typename ValueSet::value_t& v)
      : self_{std::make_shared<model<ValueSet>>(vs, v)}
    {}

    // Runtime type of the effective ValueSet.
    auto vname() const -> symbol { return self_->vname(); }

    // Extract wrapped typed Value/ValueSet pair.
    template <typename ValueSet>
    const auto& as() const {
      return dynamic_cast<const model<ValueSet>&>(*this);
    }

  private:
    class base { public:
      virtual ~base() = default;
      virtual symbol vname() const = 0;
    };

    // Type-full ValueSet/Value pair.
    template <typename ValueSet>
    class model final : public base { public:
      using value_t = typename ValueSet::value_t;
      model(const ValueSet& vs, const value_t& v)
        : valueset_(vs), value_(v) {}
      auto  vname() const override { return valueset().sname(); }
      const auto& valueset() const { return valueset_; }
      auto        value()    const { return value_; }

    private:
      const ValueSet valueset_;
      const value_t value_;
    };

    /// The wrapped value/valueset.
    std::shared_ptr<base> self_ = nullptr;
  };
  // Define dyn::label, dyn::expression, dyn::weight, etc.
  using label = value<label_tag>;
} // namespace dyn
\end{cxxcode}

There are several features of \cxx{dyn::value} that are worth being
emphasized.

First, the Value/ValueSet duality is fused: whereas in \libvcsn one
needs a ValueSet (such as \cxx{letterset}) to manipulate Values (such as
\cxx{char}), in \libdyn ``values'' are complete and self-sufficient.

Second, users of \libdyn are freed from memory management, as the \ac{api}
relies on shared pointers.  This applies for both human users, and possible
layers built on top of \libdyn (see \cref{sec:python}).

Third, we emphasize a functional-style purity: the Value and ValueSet that
\cxx{dyn::value} aggregates are immutable (\cxx{const}).

The base class, \cxx{value<Tag>::base}, is a plain class: it is not
templated.  The \cxx{value<Tag>::model} class template generates its only
derived classes.  This hierarchy provides two services only: introspection
via the \cxx{vname()} function, and type-recovery via \cxx{as()} (see
\cref{sec:dyn:algo}).

In debug mode, our type-recovery system uses
\Cxx's \ac{rtti} (\cxx{dynamic_cast}) to check for invalid conversions.
However, in practice, \cxx{static_cast} suffices: we do not use multiple
inheritance, and our introspection routines never allow a conversion with an
invalid type.

\subsection{Calling an Algorithm}
\label{sec:dyn:algo}

As a running example, consider \cxx{evaluate}, the evaluation of a word by
an automaton, which returns a weight.  Obviously, incoming arguments must be
converted from \libdyn to \libvcsn, and conversely for the result.  This is
ensured by another \cxx{evaluate} function, which we call the
\emph{static/dynamic bridge}, which uses the \cxx{as} functions for \libdyn
to \libvcsn conversion, and the \cxx{weight} (i.e., \cxx{value<weight_tag>})
constructor for the converse:
\begin{cxxcode}
namespace dyn::detail {
  template <typename Aut, typename LabelSet>
  auto evaluate(dyn::automaton aut, dyn::label lbl) -> dyn::weight
  {
    const auto& a = aut->as<Aut>();
    const auto& l = lbl->as<LabelSet>().value();
    const auto& w = ::vcsn::evaluate(a, l);
    const auto& ws = a->context()->weightset();
    return {ws, w};
  }
}
\end{cxxcode}
The bridge is actually a template: it must be parameterized by the exact
type of the wrapped automaton and wrapped label.  If \cxx{dyn::evaluate}
were to work for a single type of automaton, it would look as follows.
\begin{cxxcode}
namespace dyn {
  using automaton_t = mutable_automaton<context_t>;
  using labelset_t = labelset_t_of<context_t>;

  auto evaluate(automaton aut, label l) -> weight {
    return detail::evaluate<automaton_t, labelset_t>(aut, l);
  }
}
\end{cxxcode}
Of course it needs to accept any type of automaton and labelset.  This is
where registries come into play.

\subsection{Querying Registries}

The actual implementation of \cxx{dyn::evaluate} is:
\begin{cxxcode}
auto evaluate(automaton aut, label l) -> weight {
  auto& reg = detail::evaluate_registry();
  return reg.call(aut, l);
}
\end{cxxcode}
It uses a Meyers-style singleton \citep{meyers.04.dclp} to get
\cxx{evaluate}'s own registry:
\begin{cxxcode}
using evaluate_t = auto (automaton, label) -> weight;

static auto evaluate_registry() {
  static auto instance = registry<evaluate_t>{"evaluate"};
  return instance;
}
\end{cxxcode}
A key feature of a given bridge such as \cxx{dyn::detail::evaluate} is that
its signature does not depend on template parameters: all its instances
comply with the signature of \cxx{dyn::evaluate}, \cxx{evaluate_t}.

Each algorithm of \libdyn is associated with a single registry.  In
case of overloading, there must be one registry per overload.  For instance
\begin{cxxcode}
auto print(label  l, std::ostream& o) -> std::ostream&;
auto print(weight w, std::ostream& o) -> std::ostream&;
\end{cxxcode}
are associated with two registries: \cxx{print_label} and
\cxx{print_weight}.

The registries play the role of the tables: depending on runtime conditions,
dispatch to a single function pointer, which is specific to the type of the
arguments.  The \cxx{reg.call} invocation forwards the arguments to this
specific instance of the bridge, selected by the registry.

\subsection{Registries}
\label{sec:dyn::registry}

The registries are instances of a class template, \cxx{registry}, whose type
parameter \cxx{Fun} denotes the type of the \libdyn function such as
\cxx{evaluate_t}.

\begin{cxxcode}
template <typename Fun>
class registry { public:
  registry(const std::string& name) : name_(name) {}

  // Invoke the registered function for args.
  template <typename... Args>
  auto call(Args&&... args) {
    auto sig = vsignature(std::forward<Args>(args)...); // The signature.
    return call_sig(sig, std::forward<Args>(args)...);
  }

  // Invoke the registered function for sig, passing the args.
  template <typename... Args>
  auto call_sig(const signature& sig, Args&&... args) {
    auto bridge = get(sig);   // Get the corresponding bridge instance.
    return bridge(std::forward<Args>(args)...);           // Invoke it.
  }

  auto get(const signature& sig) { return map_.at(sig); } // Get or throw.
  auto set(const signature& sig, Fun fn) { map_[sig] = fn; }  // Register.

private:
  std::string name_;         // Function name (e.g., "evaluate").
  std::map<signature, Fun*> map_;         // Signature -> bridge.
};
\end{cxxcode}

\begin{figure}[tb]
  \centering
  \includegraphics[width=\linewidth]{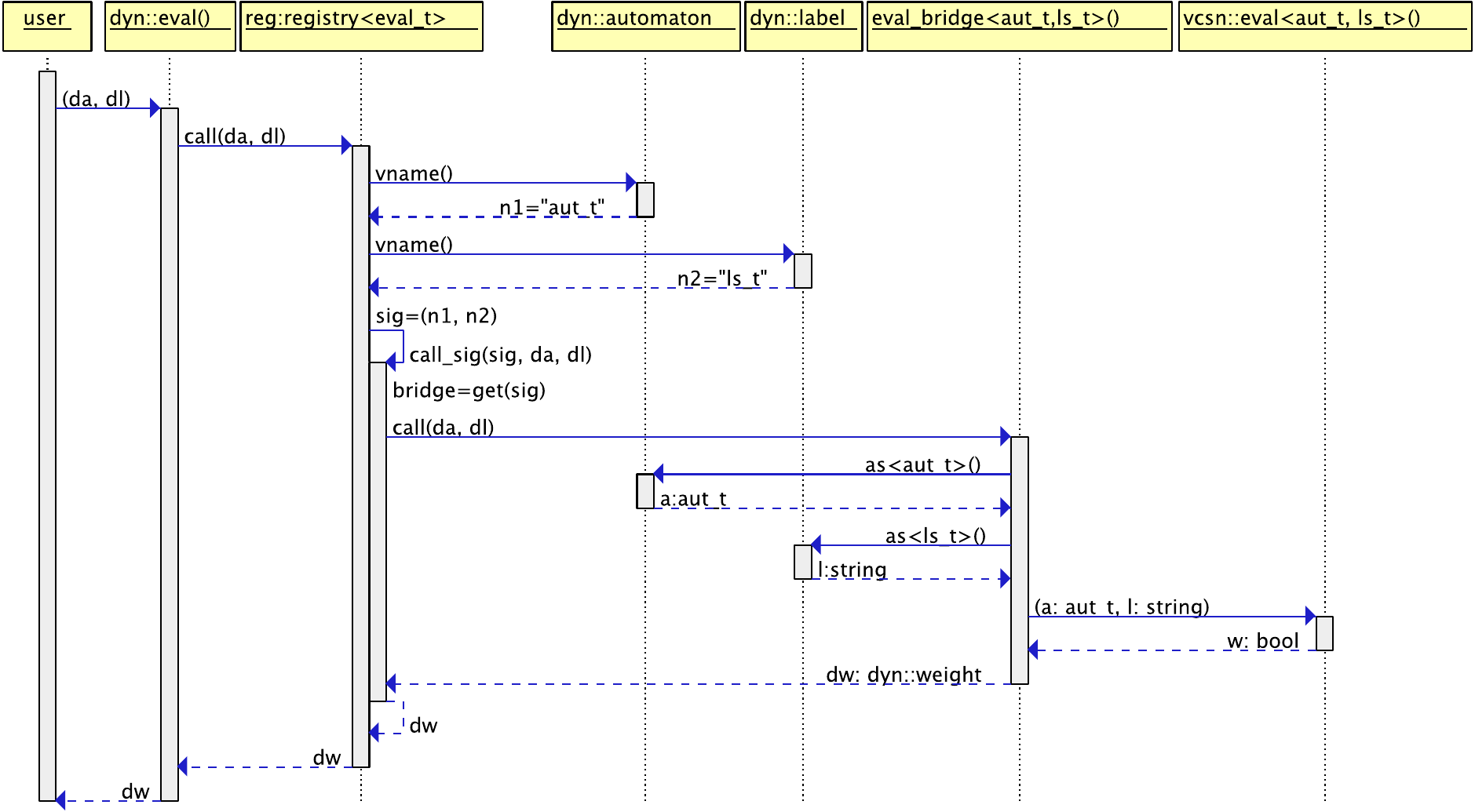}
  \caption{Sequence diagram of \cxx{dyn::evaluate(a1, l1)}, where \cxx{da}
    is a \cxx{dyn::automaton} and \cxx{dl} a \cxx{dyn::label}.}
  \label{fig:sd}
\end{figure}

To summarize, the sequence of events (see \cref{fig:sd}) when invoking
\cxx{dyn::evaluate} on an automaton of type
\cxx{mutable_automaton<letterset<char_letters>, b>} and a word of type
\cxx{wordset<char_letters>} (abbreviated \cxx{aut_t} and \cxx{ls_t}) is:
\begin{steps}
\item user calls \cxx{dyn::evaluate(da, dl)} with a \cxx{dyn::automaton} and
  a \cxx{dyn::label};
  \begin{steps}
  \item \cxx{dyn::evaluate} invokes \cxx{evaluate_registry} to obtain
    \cxx{reg}, the corresponding \cxx{registry} object;
  \item \cxx{reg.call(da, dl)} is invoked;
    \begin{steps}
    \item \cxx{call} uses \cxx{vsignature} to collect the argument vnames
      into \cxx{sig};
    \item \cxx{reg.call_sig(sig, da, dl)} is invoked;
      \begin{steps}
      \item it uses \cxx{sig} to get \cxx{bridge}, an instance of
        \cxx{detail::evaluate} parameterized by \cxx{sig};
      \item \cxx{bridge(da, dl)} is invoked;
        \begin{steps}
        \item \cxx{bridge} calls \cxx{da.as<aut_t>()} to extract \cxx{a}, (a
          \cxx{aut_t}) from \cxx{da};
        \item it extracts \cxx{l}, a \cxx{std::string}, from \cxx{dl};
        \item it calls \cxx{vcsn::evaluate(a, l)};
          \begin{steps}
          \item \cxx{vcsn::evaluate} evaluates the word;
          \item it returns a weight (a \cxx{bool});
          \end{steps}
        \item \cxx{bridge} receives this \libvcsn weight in \cxx{w};
        \item it gets \cxx{ws} (\cxx{vcsn::b}), the corresponding weightset,
          from \cxx{a};
        \item it calls \cxx{dyn::make_weight} to wrap \cxx{ws} and \cxx{w}
          into a \libdyn value;
        \item it returns this \cxx{dyn::weight};
        \end{steps}
      \item the bridge returns this \cxx{dyn::weight};
      \end{steps}
    \item \cxx{reg.call_sig} returns this \cxx{dyn::weight};
    \end{steps}
  \item \cxx{reg.call} returns this \cxx{dyn::weight};
  \end{steps}
\item \cxx{dyn::evaluate} returns this \cxx{dyn::weight} to the user.
\end{steps}

That concludes the description of the events involved in a \libdyn
invocation.  We now explain how the registries are initialized.

\subsection{Filling Registries}
\label{sec:dyn::filling}
In \vcsn, some ``contexts'' have been elected to be ``builtins'', i.e., they
are precompiled.  Examples of such contexts are $\cxx{char} \rightarrow \B$,
$\cxx{char}^* \rightarrow \B$, $\cxx{char} \rightarrow \Z$\ldots{} Each of
these contexts is compiled as a shared library.  This shared library has a
static variable \cxx{registered} whose initialization calls a function which
all registers the needed functions.  For instance, for
$\cxx{char} \rightarrow \B$:

\begin{cxxcode}
using ctx = context<letterset<char_letters>, b>;
static bool registered = register_functions<ctx>();
\end{cxxcode}
with:
\begin{cxxcode}
template <typename Ctx>
auto register_functions() {
  using aut_t = mutable_automaton<Ctx>;
  using wordset_t = wordset_t_of<Ctx>;
  // ...
  evaluate_registry().set(ssignature<aut_t, wordset_t>(),
                          detail::evaluate<aut_t, wordset_t>());
  // ...
  return true;
}
\end{cxxcode}

Classical \Cxx meta-programming techniques such as \cxx{std::enable_if} are
used to instantiate only algorithms whose preconditions are satisfied.

There is one issue, however, that this approach does not cover: the set of
possible signatures is not closed.  Consider for instance \cxx{product}: the
arguments can be of any type!  Hence, one cannot ``precompile'' all its
possible instantiations.  Code generation and dynamic shared-object loading
allow to address this issue.

\subsection{Code Generation}

Our signatures are quite heavy-weight, and look like \Cxx types.  They can
be parsed to generate \Cxx code.

Concretely, the \cxx{registry::get} function (see \cref{sec:dyn::registry})
is a little more involved:
\begin{cxxcode}
auto get(const signature& sig) {
  if (!exists(sig))
    vcsn::dyn::instantiate(name_, sig);
  return map_.at(sig);
}
\end{cxxcode}

The \cxx{instantiate} function requests the instantiation of the algorithm
named \cxx{name_} (e.g., \cxx{"evaluate"}) for the given
signature\footnote{It cannot return the function, for reasons explained
  below.}.  To do so, it generates the following piece of code:
\begin{cxxcode}
using t0_t = vcsn::mutable_automaton<vcsn::context<
               vcsn::letterset<vcsn::char_letters>, vcsn::b>>;
using t1_t = vcsn::wordset<vcsn::char_letters>;

auto sig = ssignature<t0_t, t1_t>();
auto fun = vcsn::dyn::detail::evaluate<t0_t, t1_t>;
static bool r = vcsn::dyn::detail::evaluate_registry().set(sig, fun);
\end{cxxcode}

Then \cxx{instantiate} calls the compiler and linker to produce a
shared-object, which it loads via \cxx{dlopen}.  The dynamic loader invokes
the initialization of the shared-object, i.e., initializes the \cxx{r}
variable, which registers this specific bridge in the \cxx{map_} member.
Finally \cxx{get} fetches the result in \cxx{map_}.

If the compilation failed, either for good reasons (e.g., failed
preconditions) or bad one (bugs), \cxx{instantiate} throws an exception
which is caught by the \cxx{registry} to improve the error message.  First,
it looks in the compilation log for ``static assertion failed'' messages,
which is the sign that preconditions enforced by \cxx{static_assert} were
invalidated.  If found, this information is kept; other errors are
discarded.  Then it appends the failed signature, followed by the known
ones, and finally it provides the compiler command line, to aid debugging.

If, for instance, you were to evaluate a word on an automaton labeled with
words (\cxx{evaluate} requires labels to be letters), the system would
produce the following error message:
\begin{Verbatim}[commandchars=\\\{\},formatcom={\footnotesize}]
RuntimeError: evaluate: requires a free labelset
  failed signature:
    mutable_automaton<wordset<char_letters>, b>, wordset<char_letters>
  available versions:
    mutable_automaton<letterset<char_letters>, b>, wordset<char_letters>
    mutable_automaton<letterset<char_letters>, q>, wordset<char_letters>
[...]
  failed command:
    LC_ALL=C ccache g++-4.9 -std=c++14 -fPIC '\var{base}.cc' -c -o '\var{base}.\var{pid}.o'
\end{Verbatim}
with \var{base} = \file{\var{root}/algos/evaluate/\var{sig}}, \var{root} =
\file{\$HOME/.vcsn/plugins}, and \var{sig} =
\file{mutable\_automaton<wordset<char\_letters>, b>,
  wordset<char\_letters>}.

The algorithm is compiled only once per session: the registry will not
compile it again.  To amortize the cost of compilation \emph{between}
sessions, we use \command{ccache}, a compiler cache for C and
\Cxx: only the first invocation of a non precompiled algorithm incurs the
cost of runtime compilation.

Written naively, runtime code generation, compilation, linking and loading
is exposed to race conditions: (i) two processes could be generating the
same file (in which case we'd get two copies intertwined together in the
same file, hence a compiler error), or (ii) could compile at the same time
(causing the compiler to generate garbled object file, hence a linker
error), or (iii) could link at the same time (resulting in a broken
shared-object, hence a crash).

Race conditions might seem unlikely: they require the same user to request
the same algorithm on the same signature \emph{several times, concurrently}.
However, two typical scenarios are exposed to such problems.  One is when
students use the web interface to \vcsn\footnote{\label{foot:sandbox}\vcsn
  on-line, \url{http://vcsn-sandbox.lrde.epita.fr/}.}: they all run under
the same identity, and (hopefully) all run the same requests roughly at the
same time: those from the assignment.  The other is the test suite: it runs
tests in parallel.

To address these issues we include the \var{pid}, the process identification
number, in the file names.  However, doing it blindly would void the
interest of \command{ccache}, so in step (i), once the \Cxx file is
generated, it is renamed atomically to a name independent of \var{pid}.
Steps (ii) and (iii) do produce \var{pid}'d files; afterwards the
shared-object is renamed atomically to its final name, before calling
\cxx{dlopen}.

\subsection{Customized Signatures}
\label{sec:dyn::variations}
Most algorithms in \libvcsn have a signature interfaced straightforwardly in
\libdyn:
\begin{cxxcode}
namespace dyn {
  auto evaluate(automaton aut, label l) -> weight;
  auto is_proper(automaton aut) -> bool;
}
\end{cxxcode}
However in some cases, \libdyn allows to do much better than mocking
\libvcsn.

\paragraph{Algorithms with Alternatives.}
\libvcsn features several minimization algorithms, producing automata of
different \emph{types}, see \cref{sec:algos}, so they must be different
functions.  In \libvcsn all these different automaton types are wrapped into
the unique \cxx{dyn::automaton} type, so these functions can merged into a
single one, with an additional argument to select the chosen algorithm.  We
can even do better, and provide an \cxx{"auto"} algorithm selector that
chooses the best fit, using techniques such as \cxx{std::enable_if} to
select the eligible algorithms.

\paragraph{Binary Operations.}
As a design decision, \libvcsn is picky about types.  For instance one
cannot mix weights from $\N$ with weights from $\Z$.  In \libdyn, however,
we introduce such automatic conversions.  This is done in the bridges:
\begin{cxxcode}
template <typename WS1, typename WS2>
auto sum_weight(dyn::weight lhs, dyn::weight rhs) {
  // Unwrap the dyn::weights.
  const auto& l = lhs->as<WS1>();  const auto& r = rhs->as<WS2>();
  auto ws = join(l.weightset(), r.weightset()); // Smallest common supertype.
  // Convert the input values.
  auto lr = ws.conv(l.weightset(), l.weight());
  auto rr = ws.conv(r.weightset(), r.weight());
  // Compute the result and wrap it into a dyn::weight.
  return make_weight(rs, vcsn::sum(rs, lr, rr));
}
\end{cxxcode}

\paragraph{Depending on Runtime Values.}
Applied to a multiple-tape automaton, \cxx{focus} hides all its tapes but
one.  Obviously, the resulting automaton type depends on a \emph{value} (the
tape number), not a type (\cxx{unsigned}).  Rather than turning the existing
system, based on types, into something more complex, we wrap values into
types: when \cxx{dyn::focus} is passed 2, it pretends it was actually passed
an argument whose type is \cxx{std::integral_constant<unsigned, 2>}.
However, since we need the bridges to all have the same signature,
\cxx{std::integral_constant<unsigned, 2>} cannot appear in their type.

To this end we introduce \cxx{vcsn::integral_constant}, a simple class that
stores the signature it must pretend to have
(\cxx{"std::integral_constant<unsigned, 2>"}).

To summarize \cxx{dyn::focus} is:
\begin{cxxcode}
auto focus(automaton& aut, unsigned tape) -> dyn::automaton {
  auto t = integral_constant{"std::integral_constant<unsigned, "
                             + std::to_string(tape) + '>'};
  return detail::focus_registry().call(aut, t);
}
\end{cxxcode}
and the bridge:
\begin{cxxcode}
template <typename Aut, typename Tape>
auto focus(automaton aut, integral_constant) {
  auto& a = aut->as<Aut>();
  return make_automaton(vcsn::focus<Tape::value>(a));
}
\end{cxxcode}

\paragraph{Variadic Functions.}
Because it accepts any number of automata as argument, the natural signature
for a product in \libdyn uses a vector of automata:
\begin{cxxcode}
auto product(const std::vector<automaton>& auts) -> automaton {
  return detail::product_vector_registry().call_variadic(auts);
}
\end{cxxcode}
This function invokes a different \cxx{call} function from the registry
which, instead of putting \cxx{std::vector<automaton>} in the signature,
queries each member of the vector for its signature, and collects them in a
single one, used as key:
\begin{cxxcode}
template <typename T>
auto call_variadic(const std::vector<T>& ts) {
  signature sig;
  for (const auto& t: ts)
    sig.emplace_back(vname(t));
  return call_sig(sig, ts);
}
\end{cxxcode}
This final call ensures that the template parameters of the bridge will be
the exact types of the input automata, and its (runtime) argument will be a
vector of \cxx{dyn::automaton}.  Using \Cxx{14} techniques, it unpacks the
vector into a variadic call on \libvcsn automata.
\begin{cxxcode}
template <typename... Auts>
auto product_vector(const std::vector<automaton>& auts) {
  return product_<Auts...>(auts, std::make_index_sequence<sizeof...(Auts)>{});
}

template <typename... Auts, size_t... I>
auto product_(const std::vector<automaton>& auts, std::index_sequence<I...>) {
  return make_automaton(vcsn::product(auts[I]->as<Auts>()...));
}
\end{cxxcode}

\subsection{Deserialization of Contexts}

The \libdyn library is therefore able to instantiate algorithms on the fly,
simply by asking existing objects (typically \cxx{context} instances) for
their type.  Sometimes, there is no object to start from, e.g., when loading
a file.

The \cxx{make_context} function instantiates contexts: from specification
strings such as \cxx{"lal_char(abc), b"} (or
\cxx{"context<letterset<char_letters(abc)>, b>"}, they are synonymous in
\vcsn) it builds the actual \Cxx objects they denote.  This is very
different from the previous cases where registries depend on signatures
(i.e., types) to dispatch their calls; here there is a single signature,
\cxx{(std::string)}, what changes is the \emph{value} of that string.

The \cxx{make_context} function works in two steps: code generation (to
support this specific type of contexts), and object instantiation (to
instantiate it with the appropriate runtime values, here \cxx{"abc"}).

\begin{cxxcode}
auto make_context(const std::string& spec) -> dyn::context {
  auto& reg  = detail::make_context_registry();
  auto sname = sname_normalized(spec);
  auto sig   = signature{sname};
  if (!reg.exists(sig)) vcsn::dyn::instantiate_context(sname);
  auto vname = vname_normalized(spec);
  return reg.call_sig(sig, vname);
}
\end{cxxcode}

\paragraph{Code Generation.}
The \cxx{spec} string is parsed, and pretty-printed to a canonical form,
without runtimes values (e.g., \cxx{"context<letterset<char_letters>, b>"}).
This is used to forge the signature, instead of the useless
\cxx{"(std::string)"}.  We may now query the corresponding registry to check
whether the function already exists, and request its generation if needed.

Instead of instantiating a single algorithm
(\cxx{instantiate("make_context", sig)}), we ``instantiate the context''
(\cxx{instantiate_context(sname)}): we compile the same set of signatures of
functions that were selected for the precompiled ``builtins''.  This way,
most functions are readily available when a context is first used.

\paragraph{Context Instantiation.}
The registry (see \cref{sec:dyn::registry}) is used to select the bridge of
matching \cxx{sig}, passing it \cxx{vname} as argument: the normalized
specification \emph{with} the runtime values (such as \cxx{abc}).  In turn
the bridge calls \cxx{vcsn::make_context}:
\begin{cxxcode}
template <typename Ctx>
auto make_context(const std::string& vname) -> Ctx {
  std::istringstream is{vname};  return Ctx::make(is);
}
\end{cxxcode}

All the ValueSets implement the \cxx{make} static function that allows to
deserialize them.  For instance, in the case of \cxx{context}:
\begin{cxxcode}
static auto context::make(std::istream& is) {
  eat(is, "context<");  auto ls = labelset_t::make(is);  eat(is, ',');
  auto ws = weightset_t::make(is);  eat(is, '>');
  return context{ls, ws};
}
\end{cxxcode}

\section{Discussion}
\label{sec:discussion}

In this section, we discuss the strengthes and weaknesses of our proposal.
One of the goals of \libdyn was to make easier the binding to dynamic
languages, and we expose our experience with Python.  The costs
registry-based polymorphical calls are evaluated, and we briefly discuss the
importance of \Cxx{11} in this framework.

\subsection{The Values of Value/ValueSet}
The Value/ValueSet design is the backbone of the \vcsn platform.  It
replaces the
``Element/MetaElement''~\citep{lombardy.03.ciaa,regisgianas.03.poosc} design
that was used in \vcsnu, the first incarnation of the Vaucanson project.
The purpose of Element/MetaElement is to bind together the elements (say a
\cxx{bool} value) with its MetaElement: the algebraic structure to which it
belongs (say the semiring $\B$).  The foremost advantage of the approach is
that the \Cxx operators could be overloaded: \cxx{e + f} was using the
MetaElement to get the proper implementation of the addition.  However, in
the long run, this way of structuring the library proved to be cumbersome,
and considerably hindered the development.  The library was way too complex
even for seasoned \Cxx programmers.

Value/ValueSet has a similar goal, however Values and ValueSets are never
bound together at the \libvcsn level.  In turn, this implies that operators
cannot be overloaded, and code such as
\cxx{sum = ws.add(sum, ws.mul(lhs, rhs))} --- where \cxx{ws} is the ValueSet
of \cxx{sum}, \cxx{lhs} and \cxx{rhs} --- is quite frequent in the library.
However, it turns out to be more pleasant than relying on the operators:
first it is much easier to follow the function calls (they are scoped by the
ValueSet while \Cxx operators can be defined as member or non-member
functions), second the resulting interface is more consistent: all the
operations are named, not just some of them.

In \libdyn, since Value and ValueSet are reunited, we can support operators.
Operators are offered to the end users as a nice \ac{api}, but avoided in
the implementation layers.

\medskip

ValueSets share similarities with \ac{stl}: don't put the algorithms in the
Values, keep them outside.  However, the core algorithms thanks to which a
Value implements a concept are actually grouped together by the ValueSet.

\medskip

Because ValueSets enforce consistency, in \vcsn it was surprisingly easy to
use some ValueSets in unexpected ways.  For instance, rational expressions
can be used where weights are expected, because the ExpressionSet concept is
a superconcept of WeightSet.  Conversely, because we did not follow this
guideline for our implementation of automata, we can't use them as weights.

\subsection{Benefits of \libdyn}

It is remarkable, but not unexpected, that objects and functions in \libdyn
correspond to concepts (in the \Cxx sense) of \libvcsn: WeightSets become
\cxx{dyn::weight}, the different types of automaton
(\cxx{mutable_automaton<Ctx>}, \cxx{subset_automaton<Aut>}, etc.) are mapped
to \cxx{dyn::automaton}, etc.  Even different algorithms that realize a
common specification (\cxx{minimize_moore<Aut>},
\cxx{minimize_signature<Aut>}, etc.) ---which can be seen as several
functions corresponding to a single concept--- map to a single function.

\medskip

The \libdyn \ac{api} therefore is much smaller, much simpler to use that
\libvcsn: it was designed primarily for the end users which are not
experienced \Cxx programmers.  However, the developers largely benefited
from it.  We cannot overstate how the introduction of \libdyn simplified the
development of the platform, especially wrt \ac{qa}.

Because it is very easy to create a dedicated binary for each algorithm
(e.g., \command{vcsn-evaluate}), the first generation of the test suite was
based on simple shell-scripts checking input/output combinations.  This
contrasts with the need to instantiate one program for each template
parameters set, which requires Makefile machinery, etc.

Because all the tests are on top of \libdyn, the test cases are perfectly
factored.  Testing an algorithm for a specific type of automaton always
requires some basic routines on this automaton type (e.g., input/output).
In \vcsnu these routines were compiled for each binary test case.  Thanks to
\libdyn they are shared between tests and their compilation is factored.

Before the advent of \libdyn our development cycle was the usual
``edit-compile-test'', where ``compile'' means ``compile the library'',
which may count in tens of minutes.  This is especially painful when a
single change in a header causes recompilations of components other than the
tested one.  Daredevils may recompile just the one shared-object
corresponding to the component they are testing, but that's tricky and
dangerous.  With the introduction of \libdyn, since it \emph{always}
recompiles the non-builtin \libvcsn algorithms, the development became
``edit-test'', and \libdyn deals with maintenance issues: if an algorithm
needs to be recompiled, it will be, and otherwise \command{ccache} avoids
compilation.

\medskip

Another major benefit from \libdyn is that binding to other
environments, such as Python, becomes (almost) trivial.

\subsection{Binding a Templated \Cxx Library to Python}
\label{sec:python}
Binding a templated \Cxx library such as \libvcsn in a dynamic language
(e.g., Python) has always been hard.  Tools such as
SWIG~\citep{beazley.1996.tcltk,cottom.2003.cse} are available, however, they
have a hard time coping with the whole syntax of \Cxx{} ---let alone with
its recent evolutions (e.g., \Cxx{14})--- and more importantly, cannot offer
an integration as smooth as what \libdyn provides on top of \libvcsn.

Sitting SWIG on \libdyn would be much easier, but we chose Boost.Python, a
simple means to bind \Cxx into Python~\citep{abrahams.2003.cuj}.

The \libdyn \ac{api} is an collection of free standing functions; it is not
OO to remain extensible: new function members cannot be added at runtime in
\Cxx.  Since Python supports it, our Python binding \emph{is} OO.
Evaluation in \libdyn is \cxx{evaluate(aut, word)}, in Python it is
\pyinline{aut.evaluate(word)}, or even \pyinline{aut(word)}.

Operators get their natural infix syntax: \cxx{a1.add(a2)} in \libdyn is
\pyinline{a1 + a2} in Python.  We simulate expression
templates~\citep{veldhuizen.95.c++b} and bind \pyinline{a1 & a2 & a3 & a4}
in Python into a single call to the variadic function \cxx{dyn::product}
(see \cref{sec:dyn::variations}).

\medskip

The second generation of the test suite was written as Python scripts
instead of shell-scripts calling \vcsn binaries.  This way the \vcsn
framework is loaded only once per Python script, instead of once per
\emph{component} of a test-case!  Besides, since these processes were
communicating via files or pipes, values were continuously serialized and
deserialized.

\medskip

Python was chosen especially because of IPython~\citep{perez.2007.cse},
which offers a rich graphical interactive environment.  Thanks to a very
thin additional layer, users can run commands on automata and see a
graphical rendering.  There are even dedicated widgets for interactive
edition of automata.  This environment was used successfully in three
batches of practical sessions with students.  It is available
online\cref{foot:sandbox}.

\subsection{The Cost of \libdyn Polymorphic Calls}

Extreme efficiency of the bridge between \libvcsn and \libdyn has little
importance: \vcsn was designed so that algorithms be written in \libvcsn,
not \libdyn.  Writing CPU intensive loops in \libdyn is not a realistic
scenario.  However, we performed the following measurements to evaluate its
cost.

\begin{table}[tb]
  \centering
  \newcommand{\NS}{\si{\nano\second}}
  \begin{tabular}{@{}@{~~~}r@{~~~}r@{~~~}d{2}@{~~~~~~~~~~~~}r@{~~}r@{~~~~~~~}>{\ttfamily}l>{\ttfamily}l@{}}
    \toprule
      \multicolumn{1}{c}{empty}
    & \multicolumn{1}{c}{vcsn}
    & \multicolumn{1}{l}{virt}
    & \multicolumn{1}{c}{~~~dyn~}
    & \multicolumn{1}{c}{Python~~~~}
    & Storage
    & Cast
    \\
    \cmidrule(r){1-3} \cmidrule(l){4-7}
             &          &         & 127\NS  & 371\NS & map            & dynamic \\
    0.023\NS & 0.023\NS & 1.67\NS & 110\NS  & 351\NS & map            & static  \\
             &          &         &  74\NS  & 248\NS & unordered\_map & static  \\
    \bottomrule
  \end{tabular}
  \medskip
  \caption{Average duration of one call.  We kept the fastest runs
    out of $10 \times 3$, one run making a
    loop of 1M calls. In the first row, the
    registry uses a \cxx{std::map} and \libdyn uses a \cxx{dynamic_cast}, as
    opposed to a \cxx{static_cast} in the next row.  In the last row
    we used \cxx{std::unordered_map} and \cxx{static_cast}.
    \samp{empty}: no algorithm is run at
    all, to measure the overhead of the benchmarking tool and the accuracy of
    the system clock; \samp{vcsn}: \cxx{vcsn::is_proper} is used; \samp{virt}: a
    simple \ac{oop} hierarchy calls \cxx{vcsn::is_proper} via \cxx{virtual};
    \samp{dyn}: \cxx{dyn::is_proper}; \samp{Python}:
    \pyinline{aut.is_proper()}.}
  \label{fig:is-proper}
\end{table}

The first series uses the \cxx{is_proper} algorithm: check whether an
automaton has spontaneous transitions.  What is specific about this
algorithm is that automata of the type used in the benchmark cannot have
spontaneous transitions, so the \libvcsn implementation is straightforward:
\begin{cxxcode}
template <typename Aut>
constexpr std::enable_if_t<!labelset_t_of<Aut>::has_one(), bool>
is_proper_(const Aut&) { return true; }
\end{cxxcode}
Therefore, \cxx{is_proper} is well suited to measure the cost of the dynamic
dispatch only: the algorithm itself is next to empty.  We also used our
benchmarking procedure on a empty statement.  The automaton type is
``builtin'': there is no code generation needed.

\cref{fig:is-proper} presents the results, performed on a MacBook Pro, OS X
10.9.5, Intel Core i7 2.9GHz, 8GB of RAM, using Clang 3.5, with
\samp{-DNDEBUG -O3}. The \Cxx measurements used
\cxx{std::chrono::steady_clock}, the Python measurements used
\pyinline{timeit.repeat}.  One run is a loop of one million calls, the
benching script returns the best out of three runs, and we kept the best out
of ten runs of the script.  As expected, the compiler optimizes out the
\libvcsn calls: care was taken so that in that precise case, no code is run.
An optimized \libdyn call is roughly 45x slower than a \cxx{virtual} call,
and Boost.Python/Python, in the best case, makes it 150x slower.

\begin{table}[tb]
  \centering
  \newcommand{\MS}{\si{\micro\second}}
  \begin{tabular}{ld{4}d{4}d{4}d{4}}
    \toprule
    \multicolumn{1}{c}{algorithm}
    & \multicolumn{1}{c}{vcsn}
    & \multicolumn{1}{c}{virt}
    & \multicolumn{1}{c}{dyn}
    & \multicolumn{1}{c}{Python}
    \\
    \midrule
    \cxx{thompson}      &   8.1\MS &   8.7\MS &   8.9\MS &   9.7\MS  \\
    \cxx{proper}        &  51.8\MS &  49.8\MS &  50.6\MS &  52.4\MS  \\
    \cxx{determinize}   &  19.0\MS &  19.3\MS &  19.7\MS &  20.9\MS  \\
    \cxx{minimize}      &  27.6\MS &  27.7\MS &  27.8\MS &  29.5\MS  \\
    \cxx{to_expression} &   6.6\MS &   6.9\MS &   7.7\MS &   8.9\MS  \\
    \addlinespace
    Total               & 113.1\MS & 112.4\MS & 114.7\MS & 121.5\MS \\
    \bottomrule
  \end{tabular}
  \medskip
  \caption{Average duration of one call, by keeping the fastest runs out of
    $10 \times 3$, loops of 100k calls.}
  \label{fig:kleene}
\end{table}

Highlighted this way, the differences are large.  However, they are
imperceptible in typical uses cases.  \Cref{fig:kleene} covers five typical
algorithms of automata theory, related to the Kleene theorem, on a very
moderate input: starting from the expression \texttt{[abc]*[abc]*}, build
the Thompson automaton, eliminate the spontaneous transitions
(\texttt{proper}), determinize it, minimize it, and convert the resulting
(one state!) automaton into an expression (\texttt{[abc]*}).  Results show
that even on very small inputs the cost of \libdyn calls is in the order of
the variability of the measurements (as shown by the fact that virt is
faster than \libvcsn): one call to the whole sequence lasts for
\SI{112}{\micro\second} in \libvcsn, \SI{121}{\micro\second} in Python, to
be compared to the \SI{380}{\milli\second} taken by
\command{dot}~\citep{gansner.2000.spe} to convert this one-state automaton
into SVG ---not even counting its rendering.

\subsection{Relevance of \Cxx{11}}

\Cxx{11} made the development much easier than \Cxx{98}.

\Cxx{11} plays an important role in the implementation details.  For
instance the registries use variadic templates and perfect forwarding in
\cxx{registry::call}, once to compute the signature of the actual arguments,
and then to forward these arguments to the selected bridge.  The possibility
to use \cxx{decltype} to deduce the return-type of a function call proved to
be extremely useful.

But \Cxx{11} also shows in the public \ac{api}, notably the shared pointers
in \libdyn.  Actually, \libvcsn also uses shared pointers extensively for
automata. In particular, some automata (such as \cxx{brzozowski_automaton}
or \cxx{subset_automaton}) keep track of automata they were computed from to
provide richer metadata, such as the semantics of the states.  A call to
\cxx{vcsn::strip} removes the decorator and keep only the ``naked''
automaton.  This would not have been possible had automata been plain values
rather than shared pointers.

\section{Other Works}
\label{sec:works}

Our proposal involves components that were already described in
publications.  However, most of the existing work is on language design,
rather than library design.

\subsection{Previous Work}

``Static vs. dynamic typing'', a duality that often splits programming
language communities into opponents.  Some efforts were conducted to bring
both together into a single
language~\citep{meijer.2004.oopsla,bierman.2010.ecoop}.  Our approach does
not involve changing the language we use, but rather see how some of its
features can be combined together in order to provide some of the benefits
of dynamic typing on top of a statically typed \ac{api}.

The idea of treating polymorphically \Cxx classes unrelated by inheritance
and/or having no virtual methods is not new.  This is ``external
polymorphism'', introduced twenty years go by \citet{cleeland.96.plpc}.
However it seems to be designed to cope with incompatible components, say
from different vendors.  We propose to ground the design of the library on
this idea, and to rely on introspection to be able to generate new
components.

The core of the Olena image processing project is a \Cxx highly templated
library.  They explored very soon the idea of developing a dynamic layout on
top of it~\citep{duret.00.gcse}.  However, while small scale experiments
were achieved~\citep{pouillard.06.seminar}, it never worked sufficiently
well to be integrated in the project.

Our implementation of \cxx{registry} is very similar to ``Object Factories''
\citep[Chap.~8]{alexandrescu.01.book}, an implementation of the Abstract
Factories~\citep{gamma.95.dp} tailored for \Cxx.

Our bridges implement \emph{multiple-dispatch}, ``the selection of a
function to be invoked based on the dynamic type of two or more
arguments''~\citep{pirkelbauer.2007.gpce,pirkelbauer.2010.scp}.  Compared to
the current proposals for multi-methods in
\Cxx, do not require the dispatching to be performed on members of a
classical OO hierarchy.  We also use a rich introspection system to support
code generation, and therefore a more costly dispatch mechanism.  However,
if performances were an issue, it would be interesting to rely on similar
dispatching techniques for existing functions.

\subsection{Future Works}
\paragraph{Tuning the Compiler.}
The availability of the compiler at runtime provides us with different ways
of exploiting it.  As of today, compiling a full context takes about
\SI{85}{\second} with \option{-O3 -g}, \SI{51}{\second} with \option{-O3},
\SI{32}{\second} with \option{-O0 -g}, and \SI{21}{\second} with
\option{-O0} (when \command{ccache} has a miss, otherwise consistently
\SI{0.5}{\second}).  Compiling a specific algorithm with the same options
takes \SI{30}{\second}, \SI{23}{\second}, \SI{22}{\second}, and
\SI{16}{\second}.  Therefore it would be interesting to first compile
without any optimization enabled, and then decide whether to run an
optimized compilation, possibly in background, when the algorithm is deemed
``hot''.  We can also enable profile-guided optimizations.  This is similar
to the \emph{adaptive (re)compilation}, pioneered by the Self programming
language~\citep{holzle.1991.ecoop}.

\paragraph{Better Domain Specific Syntax.}
The current syntax of the runtime type descriptors (snames) is neither
suitable for the human, nor directly exploitable as
\Cxx.  Because the user is actually exposed to it, we plan to move to a more
natural syntax, such as \cxx{char -> B}, but this syntax must be extensible
\emph{at runtime} to support user-defined ValueSets.

\section{Conclusion}
\label{sec:conclusion}

\Cxx is a widely used language to design generic and efficient libraries
crunching large data set: image/signal processing, machine learning,
computational linguistics, etc.  Interactive environments with the typical
read-eval-print loop provide a rich way to experiment with data and
algorithms.  We presented a means to build such a dynamic environment, with
an open set of types, on top of a statically typed, closed world, \Cxx
library.  To achieve this goal, we need rich runtime introspection, under
the form of complete type names.  Collections of these type names form
\dfn{signatures} which are used as keys in associative containers to get the
corresponding function template instance.  When the function is unknown, the
type names are rich enough to generate \Cxx code instantiate the missing
signature, and load it in the associative container via \cxx{dlopen}.  We
have shown how the dispatching system supports multi-methods, but also how
it permits the definition of a simple and unified \ac{api} on top of
multiple function overloads, including variadic support.  Building an
interactive environment for experiments, say using Python/IPython, becomes
extremely easy, the end user being completely shielded from template
instantiation issues.  This framework was detailed in the case of an
environment dedicated to automata, however it is applicable elsewhere,
provided a similar runtime type instropection system is implemented.

\bibliographystyle{abbrvnat}
\bibliography{%
  article,%
  share/bib/acronyms,%
  share/bib/lrde,%
  share/bib/csi,%
  share/bib/gp,%
  share/bib/comp.lang,%
  share/bib/comp.lang.c++,%
  share/bib/comp.lang.python,%
  share/bib/comp.lang.self,%
  share/bib/comp.compilers.automata%
}

\begin{thebibliography}{32}
\providecommand{\natexlab}[1]{#1}
\providecommand{\url}[1]{\texttt{#1}}
\expandafter\ifx\csname urlstyle\endcsname\relax
  \providecommand{\doi}[1]{doi: #1}\else
  \providecommand{\doi}{doi: \begingroup \urlstyle{rm}\Url}\fi

\bibitem[Abrahams and Grosse-Kunstleve(2003)]{abrahams.2003.cuj}
D.~Abrahams and R.~W. Grosse-Kunstleve.
\newblock Building hybrid systems with {B}oost.{P}ython.
\newblock \emph{C/C++ Users Journal}, 21\penalty0 (7), July 2003.

\bibitem[Alexandrescu(2001)]{alexandrescu.01.book}
A.~Alexandrescu.
\newblock \emph{Modern {C++} Design: Generic Programming and Design Patterns
  Applied}.
\newblock Addison-Wesley, 2001.

\bibitem[Allauzen et~al.(2007)Allauzen, Riley, Schalkwyk, Skut, and
  Mohri]{allauzen.07.ciaa}
C.~Allauzen, M.~Riley, J.~Schalkwyk, W.~Skut, and M.~Mohri.
\newblock Open{F}st: A general and efficient weighted finite-state transducer
  library.
\newblock In J.~Holub and J.~Zd{\'a}rek, editors, \emph{Proceedings of
  Implementation and Application of Automata, 12th International Conference
  (CIAA'07)}, volume 4783 of \emph{Lecture Notes in Computer Science}, pages
  11--23. Springer, 2007.
\newblock \url{http://www.openfst.org}.

\bibitem[Almeida et~al.(2009)Almeida, Almeida, Alves, Moreira, and
  Reis]{almeida.2009.ciaa}
A.~Almeida, M.~Almeida, J.~Alves, N.~Moreira, and R.~Reis.
\newblock {FA}do and {GUI}tar.
\newblock In S.~Maneth, editor, \emph{Proceedings of Implementation and
  Application of Automata, 14th International Conference (CIAA'09)}, volume
  5642 of \emph{Lecture Notes in Computer Science}, pages 65--74. Springer,
  2009.
\newblock ISBN 978-3-642-02978-3.

\bibitem[Beazley(1996)]{beazley.1996.tcltk}
D.~M. Beazley.
\newblock {SWIG}: An easy to use tool for integrating scripting languages with
  {C} and {C}++.
\newblock In \emph{Proceedings of the 4th Conference on USENIX Tcl/Tk Workshop,
  1996 - Volume 4}, TCLTK'96, pages 15--15, Berkeley, CA, USA, 1996. USENIX
  Association.
\newblock URL \url{http://dl.acm.org/citation.cfm?id=1267498.1267513}.

\bibitem[Bierman et~al.(2010)Bierman, Meijer, and
  Torgersen]{bierman.2010.ecoop}
G.~M. Bierman, E.~Meijer, and M.~Torgersen.
\newblock Adding dynamic types to {C}\({}^{\mbox{{\#}}}\).
\newblock In T.~D'Hondt, editor, \emph{{ECOOP} 2010 - Object-Oriented
  Programming, 24th European Conference, Maribor, Slovenia, June 21-25, 2010.
  Proceedings}, volume 6183 of \emph{Lecture Notes in Computer Science}, pages
  76--100. Springer, 2010.
\newblock ISBN 978-3-642-14106-5.
\newblock \doi{10.1007/978-3-642-14107-2}.
\newblock URL \url{http://dx.doi.org/10.1007/978-3-642-14107-2}.

\bibitem[Cleeland et~al.(1997)Cleeland, Schmidt, and
  Harrison]{cleeland.96.plpc}
C.~Cleeland, D.~C. Schmidt, and T.~Harrison.
\newblock External polymorphism --- an object structural pattern for
  transparently extending {C++} concrete data types.
\newblock In \emph{Proceedings of the 3rd Pattern Languages of Programming
  Conference}, Sept. 1997.

\bibitem[Cottom(2003)]{cottom.2003.cse}
T.~Cottom.
\newblock Using {SWIG} to bind {C}++ to {P}ython.
\newblock \emph{Computing in Science Engineering}, 5\penalty0 (2):\penalty0
  88--97, Mar 2003.
\newblock ISSN 1521-9615.

\bibitem[Demaille et~al.(2013)Demaille, Duret-Lutz, Lombardy, and
  Sakarovitch]{demaille.13.ciaa}
A.~Demaille, A.~Duret-Lutz, S.~Lombardy, and J.~Sakarovitch.
\newblock Implementation concepts in {V}aucanson 2.
\newblock In S.~Konstantinidis, editor, \emph{Proceedings of Implementation and
  Application of Automata, 18th International Conference (CIAA'13)}, volume
  7982 of \emph{Lecture Notes in Computer Science}, pages 122--133, Halifax,
  NS, Canada, July 2013. Springer.
\newblock ISBN 978-3-642-39274-0.

\bibitem[Demaille et~al.(2014)Demaille, Duret-Lutz, Lombardy, Saiu, and
  Sakarovitch]{demaille.14.ciaa}
A.~Demaille, A.~Duret-Lutz, S.~Lombardy, L.~Saiu, and J.~Sakarovitch.
\newblock A type system for weighted automata and rational expressions.
\newblock In \emph{Proceedings of Implementation and Application of Automata,
  19th International Conference (CIAA'14)}, volume 8587 of \emph{Lecture Notes
  in Computer Science}, Giessen, Germany, July 2014. Springer.

\bibitem[Duret-Lutz(2000)]{duret.00.gcse}
A.~Duret-Lutz.
\newblock Olena: a component-based platform for image processing, mixing
  generic, generative and {OO} programming.
\newblock In \emph{Proceedings of the 2nd International Symposium on Generative
  and Component-Based Software Engineering (GCSE)---Young Researchers Workshop;
  published in ``Net.ObjectDays2000''}, pages 653--659, Erfurt, Germany, Oct.
  2000.
\newblock ISBN 3-89683-932-2.

\bibitem[Gamma et~al.(1995)Gamma, Helm, Johnson, and Vlissides]{gamma.95.dp}
E.~Gamma, R.~Helm, R.~Johnson, and J.~Vlissides.
\newblock \emph{Design Patterns: {E}lements of Reusable Object-Oriented
  Software}.
\newblock {Addison-Wesley} Professional Computing Series. {Addison-Wesley}
  Publishing Company, New York, NY, 1995.

\bibitem[Gansner and North(2000)]{gansner.2000.spe}
E.~R. Gansner and S.~C. North.
\newblock An open graph visualization system and its applications to software
  engineering.
\newblock \emph{Software --- Practice and Experience}, 30\penalty0
  (11):\penalty0 1203--1233, 2000.

\bibitem[Hagberg et~al.(2008)Hagberg, Schult, and Swart]{hagberg.2008.scipy}
A.~A. Hagberg, D.~A. Schult, and P.~J. Swart.
\newblock Exploring network structure, dynamics, and function using {NetworkX}.
\newblock In \emph{Proceedings of the 7th Python in Science Conference
  (SciPy2008)}, pages 11--15, Pasadena, CA USA, Aug. 2008.

\bibitem[H\"{o}lzle et~al.(1991)H\"{o}lzle, Chambers, and
  Ungar]{holzle.1991.ecoop}
U.~H\"{o}lzle, C.~Chambers, and D.~Ungar.
\newblock Optimizing dynamically-typed object-oriented languages with
  polymorphic inline caches.
\newblock In \emph{Proceedings of the European Conference on Object-Oriented
  Programming}, ECOOP'91, pages 21--38, London, UK, UK, 1991. Springer-Verlag.
\newblock ISBN 3-540-54262-0.
\newblock URL \url{http://dl.acm.org/citation.cfm?id=646149.679193}.

\bibitem[Lombardy et~al.(2003)Lombardy, Poss, R\'egis-Gianas, and
  Sakarovitch]{lombardy.03.ciaa}
S.~Lombardy, R.~Poss, Y.~R\'egis-Gianas, and J.~Sakarovitch.
\newblock Introducing {V}aucanson.
\newblock In O.~H. Ibarra and Z.~Dang, editors, \emph{Proceedings of
  Implementation and Application of Automata, 8th International Conference
  (CIAA'03)}, volume 2759 of \emph{Lecture Notes in Computer Science}, pages
  96--107, Santa Barbara, CA, USA, July 2003. Springer.

\bibitem[Meijer and Drayton(2004)]{meijer.2004.oopsla}
E.~Meijer and P.~Drayton.
\newblock Static typing where possible, dynamic typing when needed: The end of
  the cold war between programming languages.
\newblock In \emph{OOPSLA'04 Workshop on Revival of Dynamic Languages}, 2004.

\bibitem[Meyers and Alexandrescu(2004)]{meyers.04.dclp}
S.~Meyers and A.~Alexandrescu.
\newblock {C++ and the Perils of Double-Checked Locking}.
\newblock \emph{Dr. Dobb's Journal}, July 2004.
\newblock
  \url{http://www.aristeia.com/Papers/DDJ\_Jul\_Aug\_2004\_revised.pdf}.

\bibitem[Myers(1995)]{myers.95.cppr}
N.~C. Myers.
\newblock Traits: a new and useful template technique.
\newblock \emph{{C++} Report}, 7\penalty0 (5):\penalty0 32--35, June 1995.
\newblock \url{http://www.cantrip.org/traits.html}.

\bibitem[Peixoto()]{peixoto.2014.gtp}
T.~P. Peixoto.
\newblock Graph-tool performance comparison.
\newblock \url{http://graph-tool.skewed.de/performance}, Retrieved on
  2014-12-03.

\bibitem[Peixoto(2014)]{peixoto.2014.graph-tool}
T.~P. Peixoto.
\newblock The graph-tool {P}ython library.
\newblock \emph{figshare}, 2014.
\newblock URL \url{http://figshare.com/articles/graph_tool/1164194}.

\bibitem[P\'erez and Granger(2007)]{perez.2007.cse}
F.~P\'erez and B.~E. Granger.
\newblock {IP}ython: a system for interactive scientific computing.
\newblock \emph{Computing in Science and Engineering}, 9\penalty0 (3):\penalty0
  21--29, May 2007.
\newblock ISSN 1521-9615.
\newblock \doi{10.1109/MCSE.2007.53}.
\newblock URL \url{http://ipython.org}.

\bibitem[Pirkelbauer et~al.(2007)Pirkelbauer, Solodkyy, and
  Stroustrup]{pirkelbauer.2007.gpce}
P.~Pirkelbauer, Y.~Solodkyy, and B.~Stroustrup.
\newblock Open multi-methods for {C}++.
\newblock In \emph{Proceedings of the 6th International Conference on
  Generative Programming and Component Engineering}, GPCE'07, pages 123--134,
  New York, NY, USA, 2007. ACM.
\newblock ISBN 978-1-59593-855-8.
\newblock URL \url{http://doi.acm.org/10.1145/1289971.1289993}.

\bibitem[Pirkelbauer et~al.(2010)Pirkelbauer, Solodkyy, and
  Stroustrup]{pirkelbauer.2010.scp}
P.~Pirkelbauer, Y.~Solodkyy, and B.~Stroustrup.
\newblock Design and evaluation of c++ open multi-methods.
\newblock \emph{Sci. Comput. Program.}, 75\penalty0 (7):\penalty0 638--667,
  July 2010.
\newblock ISSN 0167-6423.
\newblock \doi{10.1016/j.scico.2009.06.002}.
\newblock URL \url{http://dx.doi.org/10.1016/j.scico.2009.06.002}.

\bibitem[Pouillard and Thivolle(2006)]{pouillard.06.seminar}
N.~Pouillard and D.~Thivolle.
\newblock Dynamization of {C}++ static libraries.
\newblock Technical Report 0602, EPITA Research and Development Laboratory
  (LRDE), 2006.

\bibitem[R\'egis-Gianas and Poss(2003)]{regisgianas.03.poosc}
Y.~R\'egis-Gianas and R.~Poss.
\newblock On orthogonal specialization in {C++}: dealing with efficiency and
  algebraic abstraction in {V}aucanson.
\newblock In J.~Striegnitz and K.~Davis, editors, \emph{Proceedings of the
  Parallel/High-performance Object-Oriented Scientific Computing (POOSC; in
  conjunction with ECOOP)}, number FZJ-ZAM-IB-2003-09 in John von Neumann
  Institute for Computing (NIC), pages 71--82, Darmstadt, Germany, July 2003.

\bibitem[Sakarovitch(2009)]{sakarovitch.09.eat}
J.~Sakarovitch.
\newblock \emph{Elements of Automata Theory}.
\newblock Cambridge University Press, 2009.
\newblock Corrected English translation of \emph{\'El\'ements de th\'eorie des
  automates}, Vuibert, 2003.

\bibitem[Siek et~al.(2001)Siek, Lee, and Lumsdaine]{siek.01.bgl}
J.~G. Siek, L.-Q. Lee, and A.~Lumsdaine.
\newblock \emph{The {Boost Graph Library}: User Guide and Reference Manual}.
\newblock {C++} In-Depth Series. Addison Wesley Professional, 1st edition, Dec.
  2001.
\newblock ISBN 0-201-72914-8.

\bibitem[Sutton(2014)]{sutton.2014.n4040}
A.~Sutton.
\newblock Working draft, {C}++ extensions for concepts.
\newblock Technical Report N4040, May 2014.

\bibitem[Sutton et~al.(2013)Sutton, Stroustrup, and
  {Dos~Reis}]{sutton.2013.3701}
A.~Sutton, B.~Stroustrup, and G.~{Dos~Reis}.
\newblock Concepts lite.
\newblock Technical Report N3701, Texas A\&M University, June 2013.
\newblock
  \url{http://www.open-std.org/jtc1/sc22/wg21/docs/papers/2013/n3701.pdf}.

\bibitem[{The Boost Project}()]{boost.www}
{The Boost Project}.
\newblock {Boost} {C++} libraries.
\newblock \url{http://www.boost.org/}.

\bibitem[Veldhuizen(1995)]{veldhuizen.95.c++b}
T.~L. Veldhuizen.
\newblock Expression templates.
\newblock \emph{C++ Report}, 7\penalty0 (5):\penalty0 26--31, June 1995.
\newblock ISSN 1040-6042.
\newblock Reprinted in C++ Gems, ed. Stanley Lippman.

\end{thebibliography}
\end{document}